\documentclass[superscriptaddress,showpacs,aps,twocolumn,prb,floatfix]{revtex4}
\usepackage{bm,amsmath,amssymb}
\usepackage{graphicx}
\usepackage{color}

\begin{document}

\title{Entangled end states with fractionalized spin projection in a 
time-reversal-invariant topological superconducting wire}
\author{Armando A. Aligia}
\affiliation{Centro At\'omico Bariloche and Instituto Balseiro, CNEA, 8400 S. C. de
Bariloche, Argentina}
\author{Liliana Arrachea}
\affiliation{International Center for Advanced Studies, ECyT-UNSAM, Campus Miguelete, 25
de Mayo y Francia, 1650 Buenos Aires, Argentina}

\begin{abstract}
We study the ground state and low-energy subgap excitations of a
finite wire of a time-reversal-invariant topological superconductor (TRITOPS) with spin-orbit coupling. We
solve the problem analytically for a long chain of a specific one-dimensional lattice model 
in the electron-hole symmetric configuration and numerically for other cases of the same model. 
We present results for the spin density of excitations in long chains with 
an odd number of particles. The total spin projection along the axis of the spin-orbit coupling $S_z= \pm 1/2$
is distributed with fractions $\pm 1/4$  
localized at both ends, and shows even-odd alternation along the sites of the chain. 
We calculate the localization length of these excitations and find that it can be
well approximated by a simple analytical expression. We show that the energy $E$ of the lowest 
subgap excitations of the finite chain defines
tunneling and entanglement between end states.
We discuss the effect of a Zeeman coupling $\Delta_Z$ on one of the ends of the chain only. 
For $\Delta_Z<E$,  the energy difference 
of excitations with opposite spin orientation is  $\Delta_Z/2$,  consistent with a spin projection $\pm 1/4$.
We argue that these physical features are not model dependent and can be experimentally observed in
TRITOPS wires under appropriate conditions.
\end{abstract}

\maketitle

\section{Introduction}

\label{intro}

Topological materials including topological superconductors (TS) are a subject of
great interest recently in condensed matter physics.
This field of research had a burst after  the observation by Kitaev that a one-dimensional model of 
spinless fermions with $p$-wave BCS pairing has a topological phase with 
zero-energy subgap excitations  that  are described by Majorana fermions.\cite{kitaev-model} 
The non-abelian statistics obeyed by these quasiparticles is an appealing property 
for implementing quantum computing protocols. Since then, several proposals for realizing this phase in concrete physical systems were formulated. In particular,   
quantum superconducting  wires with spin-orbit coupling and magnetic field,\cite{wires1,wires2, wires-exp1,wires-exp2,wires-exp3,wires-exp4,wires-exp5}  
edge states of the quantum spin Hall state in proximity to superconductors and in contact to magnetic 
moments,\cite{fu} and Shiba states induced by magnetic adatoms on superconducting substrates.\cite{nadj} 
All these mechanisms to generate the topological superconducting phase
contain ingredients breaking time-reversal symmetry.  

\begin{figure}[h]
\begin{center}
\includegraphics[width=\columnwidth]{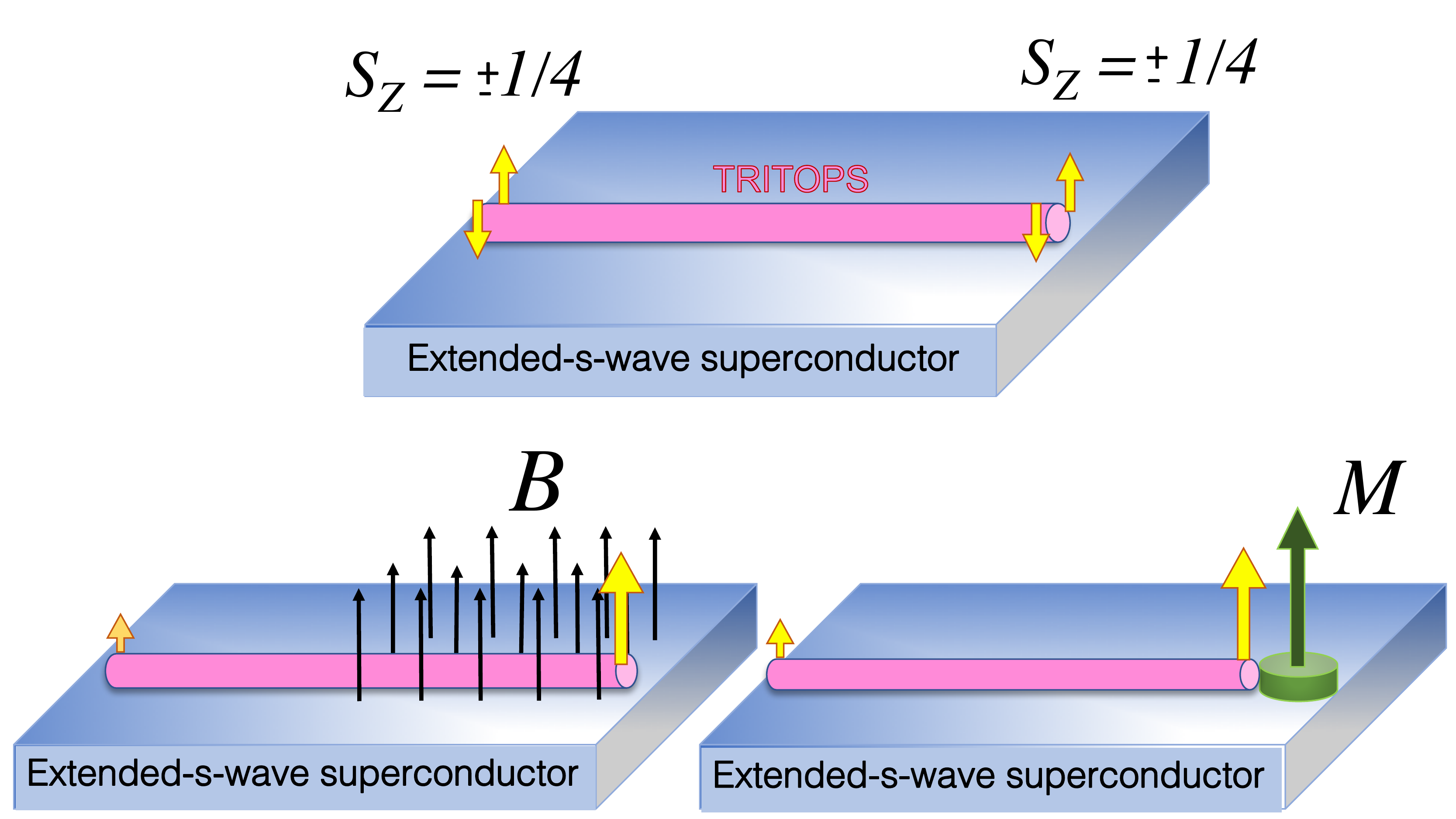}
\end{center}
\caption{(Color online) Sketch of the setup. Top: Excited subgap states with total spin projection $S_z= 1/2$ 
of a TRITOPS wire with spin-orbit coupling, where superconductivity is induced by proximity
to a macroscopic superconductor with extended s-wave pairing. Spin projection is fractionalized  with  
$S_e^z= 1/4$ localized at the ends of the wire.
Bottom: a weak magnetic field  $B$ in the direction of the spin-orbit coupling  is applied at the right
side of the setup, inducing polarization of the subgap mode (Left). 
Alternatively, one of the subgap modes interacts with a magnetic island with magnetic moment $M$ (Right).}
\label{fig0}
\end{figure}

In contrast, there is another family of TS, the time-reversal-invariant topological superconductors 
(TRITOPS) where the zero-mode edge excitations 
appear in Kramers pairs.\cite{dumi1,fanz,kesel,haim,yuval,klino,tritops-bt,tritops-ort,chung,yaco} 
This property has interesting implications which can be relevant for their 
detection and manipulation. \cite{scha,tritops-ber,cam,jose1,para,bs} For a recent review on proposals to realize the TRITOPS phase, see Ref. \onlinecite{review}.
In particular, Zhang \textit{et al.} 
\cite{fanz} proposed to engineer one- and two-dimensional
TRITOPS via proximity
effect between nodeless extended $s$-wave iron-based superconductors and
semiconducting systems with large Rashba spin-orbit interactions.  A sketch of a one-dimensional (1D) setup is shown in Fig. \ref{fig0}. 
At each end of a
long TRITOPS wire, there is a Kramers pair of Majorana edge states at zero
energy. For a finite wire, there is a mixing of the end states and the four
fermions of zero energy split in two pairs with energy $\pm E$. One of the interesting properties of 
this family is the fact that subgap excitations were argued to have fractional spin projection 
along the direction of the spin-orbit coupling $z$.\cite{kesel} 
This has consequences 
in the physical behavior of these systems when put in contact to magnetic systems. 
An example is the quench of the $0-\pi$ transition in the Josephson  current
of a quantum dot embedded in a TRITOPS junction.\cite{cam}

In this work we study these low-energy end states of a finite chain. By using a method 
presented recently by Alase \textit{et al.} \cite{alas1,alas2} we analytically calculate the 
zero-energy eigenstates of an infinite chain of the model introduced by Zhang \textit{et al.}  
in Ref. \onlinecite{fanz} in the
particle-hole symmetric configuration of the normal system (which means that the chemical potential $\mu=0$). 
This corresponds to the explicit  solution of the  Kramers pairs of Majorana edge states. 
We calculate the localization length of the low-energy excitations for arbitrary $\mu$.
We also find analytical explicit expressions for these states in the case of finite chains with $\mu=0$ and complement
our study with some numerical results for  other values of $\mu$.  
In finite chains with an odd number of particles, there is an effective tunneling which entangles the end states. 
This stabilizes a ground state in which the spin projection at each end is $S_e^z= \pm 1/4$ 
[$e=L$ (right) or $e=R$ (right)]. Instead, for systems with an even number of particles, $S_e^z=0$. 
We show that the parameter characterizing the tunneling is, 
precisely, the energy $E$ of the lowest subgap excitations. We also calculate the distribution of 
$S_z$ along the chain.
In addition, we analyze the response  to Zeeman coupling $\Delta_Z$
induced by a weak magnetic field or an Ising coupling with a magnetic moment,  
acting on one of the edges of the wire, as sketched in Fig. \ref{fig0}.
We show that 
in short enough chains where the excitations at both ends are entangled,
the fractionalization of $S_z$ manifests itself in a Zeeman splitting of half 
the amplitude of the usual one for spin 1/2.
The condition to observe this fractional Zeeman response 
is $\Delta_Z < E$. 
For long chains with $\Delta_Z \gg E$, the spin projection at each end remains $S_e^z= \pm 1/4$
depending on the total spin projection $S_z= \pm 1/2$ for odd number of particles,
while it evolves to $S_L^z= - S_R^z= \pm 1/4$ for even number of particles.
Importantly, although we solve a specific model, the physical behavior related to the 
distribution of $S_z$ and response is  generic of any TRITOPS wire. 

The finite energy of these odd parity states might be detected
in experiments where capacitance effects permit to
control the charge in small superconducting islands.\cite{lafar}
In these systems also the chemical potential $\mu$ can be controlled by 
a gate voltage. The Zeeman spitting can also be detected by  scanning tunneling spectroscopy experiments 
akin to those performed to observe Shiba states induced by magnetic adatoms in superconducting 
substrates.\cite{yazdani,pascual,franke}.
Finally also microwave excitations \cite{tosi,hays} in a finite chain with odd number of particles
might detect the anomalous (half) Zeeman splitting of the low-energy excitations.

The paper is organized as follows. In Sec, \ref{model} the model is described.
In Sec. \ref{alasec} we present the approach to analytically diagonalize it  for $\mu=0$. 
In Sec. \ref{loca} we show the dependence on the localization length with $\mu$
and compare with a simple analytical approximation.
Sec. \ref{spin} contains analytical and numerical results for the spin projection $S_i^z$ at
each site $i$ for a finite wire with odd number of particles. 
In Sec. \ref{field} we calculate the effect of a magnetic field at one end of the chain.
In Sec. \ref{sum} we present a summary and a brief discussion.

\section{Model}

\label{model}

The TRITOPS chain is described by the Hamiltonian proposed in Ref. \onlinecite{fanz}

\begin{eqnarray}
H &=&\sum_{j=1}^{L}\sum_{\sigma }[-tc_{j+1\sigma }^{\dagger }c_{j\sigma
}-\mu \;c_{j\sigma }^{\dagger }c_{j\sigma }  \notag \\
&&+i\lambda _{\sigma }c_{j+1\sigma }^{\dagger }c_{j\sigma }+(\Delta _{\sigma}e^{i\phi }c_{j+1\sigma }^{\dagger }c_{j\overline{\sigma }}^{\dagger }+%
\mathrm{H.c.})],
\label{ham}
\end{eqnarray}%
where $\lambda _{\uparrow ,\downarrow }=\pm \lambda $, $\Delta _{\uparrow
,\downarrow }=\pm \Delta $ and $\overline{\uparrow }=\downarrow ,\;\overline{%
\downarrow }=\uparrow $. The first term corresponds to nearest-neighbor
hopping, $\mu $ is the chemical potential, and $\lambda $ and $\Delta $ are
the strengths of Rashba spin-orbit coupling and extended s-wave pairing,
respectively. 

For completeness, we include in the Hamiltonian 
the phase $\phi $, which   is important when the chain is coupled in a
Josephson circuit, although it does not play an important role in the behavior of 
the spin excitations. \cite{fanz,cam,jose1,jose2}  For $\phi =0$, the Hamiltonian is invariant under time
reversal symmetry. In addition, in absence of superconductivity ($\Delta =0$), 
for $\mu =0$, the Hamiltonian is invariant under the electron hole
transformation $c_{j\sigma }^{\dagger }\rightarrow (-1)^{j}c_{j\sigma }$. 

While $H$ supports topological and nontopological phases, in our work we are
interested in the topological phase that takes place for $|\mu |< 2 |\lambda| $.
In Ref.  \onlinecite{fanz} a local s-wave pairing $\Delta_0$ was also considered. 
Since the topological phase exists for dominant nearest-neighbor pairing ($|\Delta_0| < |\Delta|$), 
for simplicity we focus on the case   $\Delta_0=0$. 
In this phase when the number of sites $L\rightarrow \infty $, there is a
Kramers pair of Majorana fermions at each end with energy $E=0$. For a
finite chain the end states mix as described in the following Sections.

\section{Diagonalization of the chain with arbitrary boundary conditions}

\label{alasec}

In this Section, we discuss the application of the method presented in Refs. 
\onlinecite{alas1,alas2} by Alase {\it et al.} to diagonalize a 1D non-interacting homogeneous 
Hamiltonian with arbitrary boundary conditions
 Then, we study the particular case of an electron-hole
symmetric band ($\mu =0$) for which an analytical result for the zero-energy
eigenstates for $L \rightarrow \infty $ is derived. Finally we also obtain
analytically the low-energy eigenstates for $\mu =0$ and a long finite chain.
Those readers who are not interested in the derivation can skip this section,
and go directly to the analytical results for $\mu=0$: 
Eq. (\ref{pe}) and the following for the zero-energy modes of the infinite chain,
and Eq. (\ref{ene}) and the following for the low-energy modes of the finite chain.

\subsection{Formalism for the general case}

Here we we discuss the application of the method of Alase {\it et al.} to the model of Eq. (\ref{ham}).
For those readers who are more familiar with Nambu notation, an alternative 
version of the procedure is presented in Appendix A.

To simplify the use of the method of Alase {\it et al.},\cite{alas1,alas2} 
it is convenient to map the model with $M$
spin-orbitals $\alpha $ ($M=2L$ in our case) to one expressed in terms of $2M$ kets
associated with the annihilation ($a$) and creation ($c$) operators

\begin{equation}
c_{\alpha }\leftrightarrow |\alpha a\rangle \text{, }c_{\alpha }^{\dagger}
\leftrightarrow |\alpha c\rangle .  \label{map}
\end{equation}%
The ensuing Hamiltonian is 

\begin{eqnarray}
 \tilde{H} &=& \sum_{\beta \alpha} \left( A_{\beta \alpha} |\beta a \rangle \langle \alpha a | + B_{\beta \alpha} |\beta c \rangle \langle \alpha a | \nonumber  \right.\\
& &  \left. -
 \overline{A}_{\beta \alpha} |\beta c \rangle \langle \alpha c | + \overline{B}_{\beta \alpha} |\beta a \rangle \langle \alpha c | \right),
 \end{eqnarray}
 with $\overline{A}_{\beta \alpha}={A}^*_{\beta \alpha}$ and $\overline{B}_{\beta \alpha}={B}^*_{\beta \alpha}$. These matrix elements are 
  defined from the equations

\begin{equation}
\lbrack c_{\alpha },H]=\sum_{\beta }(A_{\beta \alpha }c_{\beta }+B_{\beta
\alpha }c^{\dagger}_{\beta }).  \label{com}
\end{equation}%
Hence

\begin{eqnarray}
\tilde{H}|\alpha a\rangle &=&\sum_{\beta }(A_{\beta \alpha }|\beta a\rangle
+B_{\beta \alpha }|\beta c\rangle ),  \notag \\
\tilde{H}|\alpha c\rangle &=&-\sum_{\beta }(\bar{A}_{\beta \alpha }|\beta
c\rangle +\bar{B}_{\beta \alpha }|\beta d\rangle ).  \label{htilde}
\end{eqnarray}

Thus, we see that  solving Eq. (\ref{htilde}) is equivalent to solving Eq. (\ref{com}).

In this  notation we define  projection operators over bulk ($P_{B}$) 
and boundary ($P_{\text{bou}}$) states with $P_{B}+P_{\text{bou}}=1 $. 
The projector $P_{B}$ is over all those sites in which all the hopping terms
are contained in the chain. In our case

\begin{eqnarray}
P_{B} &=&\sum_{j=2}^{L-1}P_{j},  \notag \\
P_{\text{bou}} &=&P_{1}+P_{L},  \notag \\
P_{j} &=&\sum_{\sigma }(|j\sigma a\rangle \langle j\sigma a|+|j\sigma
c\rangle \langle j\sigma c|).  \label{proj}
\end{eqnarray}%
Following again Alase et al., \cite{alas1,alas2} we construct generalized Bloch functions which permit to solve the bulk
eigenvalue problem $P_{B}\tilde{H}|e\rangle =EP_{B}|e\rangle $
  for certain
roots $z(E)$. Finally the equation $P_{\text{bou}}\tilde{H}|e\rangle =P_{%
\text{bou}}E|e\rangle $ determines the allowed energies $E$ and the
eigenstates $|e\rangle $. Specifically, for our problem the four generalized Bloch
states can be written as

\begin{equation}
|z\sigma o\rangle =w_{z}\sum_{j=1}^{L}z^{j-1}|j\sigma o\rangle,
\label{bloch}
\end{equation}
where $o=a$ or $c$, $\sigma=\uparrow$ or $\downarrow$, $w_{z}$ is the coefficient of $|j\sigma o\rangle $ and $%
z $ is a complex number to be determined later.

Using Eqs.  (\ref{ham}), (\ref{com}), (\ref{htilde}), (\ref{proj}) and (\ref{bloch}) and some algebra we get

\begin{eqnarray}
P_{B}\tilde{H}|z \uparrow a\rangle &=&[-\mu -t\left( z+\frac{1}{z}\right) 
\notag \\
&+&i\lambda \left( z-\frac{1}{z}\right) ]P_{B}|z \uparrow a\rangle  \notag \\
&+&\Delta e^{i\phi }\left( z+\frac{1}{z}\right) P_{B}|z \downarrow c\rangle ,
\notag \\
P_{B}\tilde{H}|z \downarrow c\rangle &=&[\mu +t\left( z+\frac{1}{z}\right) 
\notag \\
&-&i\lambda \left( z-\frac{1}{z}\right) ]P_{B}|z \downarrow c\rangle  \notag
\\
&+&\Delta e^{-i\phi }\left( z+\frac{1}{z}\right) P_{B}|z \uparrow a\rangle .
\label{pbh}
\end{eqnarray}%
Similar equations are obtained interchanging $\uparrow $ and $\downarrow $
and simultaneously changing the sign of both, $\lambda $ and $\Delta $.

Using Eqs.  (\ref{pbh}), and the ansatz $|e\rangle =u|z\uparrow a\rangle
+v|z\downarrow c\rangle $, the bulk eigenvalue problem $P_{B}(\tilde{H}-E)|e\rangle =0$ 
takes the form

\begin{eqnarray}
\left( 
\begin{array}{cc}
a(z)-E \text{  }b(z) e^{-i\phi } \\ 
b(z) e^{i\phi }  \text{  }-a(z)-E
\end{array}
\right) \left( 
\begin{array}{c}
u(z) \\ 
v(z)
\end{array}
\right) =0,  \label{eig}
\end{eqnarray}
where
\begin{eqnarray}
a(z) &=&-\mu -t\left( z+\frac{1}{z}\right) +i\lambda \left( z-\frac{1}{z}%
\right) ,  \notag \\
b(z) &=&\Delta \left( z+\frac{1}{z}\right) .  \label{eigb}
\end{eqnarray}
Vanishing of the determinant of the matrix implies

\begin{eqnarray}
E^{2} &=&\mu ^{2}+\left( t^{2}+\Delta ^{2}\right) \left( z+\frac{1}{z}%
\right) ^{2}-\lambda ^{2}\left( z-\frac{1}{z}\right) ^{2}  \notag \\
&&+2\mu t\left( z+\frac{1}{z}\right)  \notag \\
&&-2i\lambda \left[ \mu \left( z-\frac{1}{z}\right) +t\left( z^{2}
-\frac{1}{z^{2}}\right) \right] .  
\label{e2}
\end{eqnarray}
For each energy $E$, there are four solutions $z_{k}(E)$ of this equation,
which lead to four eigenstates $|e_{k}\rangle $ of the bulk equation with $u_{k}>0$ and
\begin{equation}
\frac{v_{k}}{u_{k}}=\frac{e^{i\phi }\left( E-a_{k}\right) }{b_{k}}.
\label{vsu}
\end{equation}
Here we have introduced the notation $a_k \equiv a(z_k)$, $b_k \equiv b(z_k)$,
$u_{k}\equiv u_{\sigma}(z_k)$ and  $v_{k}\equiv v_{\sigma}(z_k)$.

The solution of the full eigenvalue equation $(\tilde{H}-E)|f\rangle =0$ has
the form $|f\rangle =\sum \alpha _{k}|e_{k}\rangle $. The coefficients $%
\alpha _{k}$ and the energy $E$ are determined by the boundary equations $P_{%
\text{bou}}(\tilde{H}-E)|f\rangle =0$. For our problem $P_{1}(\tilde{H}%
-E)|f\rangle =0$ and $P_{L}(\tilde{H}-E)|f\rangle =0$ implies

\begin{eqnarray}
&&\sum_{k=1}^{4}\left[ \left( a_{k}^{1}-E\right) u_{k}+b_{k}^{1}e^{-i\phi
}v_{k}\right] \alpha _{k} =0,  \notag \\
&&\sum_{k=1}^{4}\left[ u_{k}b_{k}^{1}e^{i\phi }-\left( a_{k}^{1}+E\right)
v_{k}\right] \alpha _{k} =0,  \notag \\
&&\sum_{k=1}^{4}\left[ \left( a_{k}^{L}-E\right) u_{k}+b_{k}^{L}e^{-i\phi
}v_{k}\right] \alpha _{k} =0,  \notag \\
&&\sum_{k=1}^{4}\left[ b_{k}^{L}e^{i\phi }u_{k}-\left( a_{k}^{L}+E\right)
v_{k}\right] \alpha_{k} =0,   \label{ep1}
\end{eqnarray}%
with the definitions
\begin{eqnarray}
&&a_{k}^{1} =w_{k}\left( -\mu -tz_{k}+i\lambda z_{k}\right) ,  \notag \\
&&a_{k}^{L} =w_{k}\left( -\mu -\frac{t}{z_{k}}-i\frac{\lambda}{z_{k}}\right) z_{k}^{L-1} ,  \notag \\
&&b_{k}^{1} =w_{k}\Delta z_{k}, \;\;\;\;\;\;\;\;\; \;\;\;\;\;\;b_{k}^{L} =w_{k}\Delta z_{k}^{L-2}.  \label{ep2}
\end{eqnarray}
Interchanging $\uparrow $ and $\downarrow $ and simultaneously changing the
sign of both, $\lambda $ and $\Delta $, or directly applying the time
reversal operator for $\phi =0$, eigenstates (degenerate with the
previous ones) which involve linear combinations of the form $u|z\downarrow
a\rangle +v|z\uparrow c\rangle $ are obtained.

\subsection{Solution for $\mu =0$, $L\rightarrow \infty $, $E\rightarrow 0$}
\label{linf}

In general, for finite wires, the chemical potential $\mu $ can be adjusted
using a gate voltage. For $\mu =0$, Eq.  (\ref{e2}) can be solved
analytically. In fact, for $\mu =0$ the    odd powers of $z$ in Eq.  (\ref{e2}) disappear and the ensuing equation is  quadratic
in $z^{2}$.  
This value of the chemical potential is within the topological phase, which is characterized by
Kramers pairs of Majorana zero modes localized at the ends of the chain. 
These states are exactly at $E=0$ only in the limit of
an infinitely long chain, where they are completely decoupled. 
We discuss this limit here since this solution sheds light on the structure and properties of these
states. In particular for $t\rightarrow 0$, $|\Delta |=|\lambda |$ a very simple solution is found,
which involves operators at sites 1 and $L$ only.
We postpone the discussion of the effect of the finite length of the chain for the next subsection.

Note that for any $\mu$, Eq.  (\ref{e2}) is invariant under the simultaneous change of sign of 
$\lambda$ and  $z\leftrightarrow 1/z$. 
In addition,  if $z$ is a solution of (\ref{e2}), its complex conjugate 
$\bar{z}$ is a solution of the same equation for the opposite 
sign of $\lambda$. Combining both properties one realizes that 
if $z$ is a solution of Eq. (\ref{e2}),  $1/\bar{z}$ is also a solution.
For $\mu=0$, if $z$ is a solution,  $-z$ is also a solution.
This implies that knowing one solution of Eq.  (\ref{e2}), 
which we call $z_{1}$, the others are related to it as follows

\begin{equation}
z_{2}=-z_{1}\text{, }z_{3}=\frac{1}{\overline{z}_{1}}\text{, }z_{4}=-z_{3}\text{. 
}  \label{zs}
\end{equation}

For $E=0$ two solutions have $|z|<1$ and the other two $|z|>1$. Choosing $%
z_{1}$ as one of the  solutions satisfying $|z_1|<1$,  we obtain after some algebra

\begin{equation}
z_{1}^{2}=z_{2}^{2}=-\frac{t^{2}+\left( |\Delta |-|\lambda |\right) ^{2}}{%
\Delta ^{2}+\left( t-i\lambda \right) ^{2}}.  \label{z2}
\end{equation}%
Then, the amplitude of $|j\sigma o\rangle $ in the states $|e_{k}\rangle $
with $k=1,2$ ($k=3,4$) decrease (increase) exponentially as $j$ increases.
In the limit where $L \rightarrow \infty $ we can separate the states localized at
each end. For the left one (low $j$) only $k=1,2$ matter. Using Eq.  (\ref{vsu}) we obtain

\begin{equation}
\frac{v_{1}}{u_{1}}=\frac{v_{2}}{u_{2}}=p,  \label{vsu1}
\end{equation}%
where we define

\begin{equation}
p=ie^{i\phi }\text{sign}(\Delta \lambda ).  \label{pe}
\end{equation}%
We also define $w_{k}=1/u_{k}$ and normalize the states at the end of the calculation. With this choice, 
the last two Eqs.  (\ref{ep1}) lead to $a_{2}^{1}=-a_{1}^{1}$, $b_{2}^{1}=-b_{1}^{1}$, 
and the first two Eqs.  (\ref{ep1}) lead to $\alpha_{2}=\alpha _{1}$. In terms of the original 
fermionic operators, the normalized
solution for the zero-energy $\uparrow$ spin excitation localized at the left end reads

\begin{eqnarray}
\gamma _{\uparrow } &=&N\sum_{n=0}^{\infty }z_{1}^{2n}\left( c_{2n+1\uparrow
}+pc_{2n+1\downarrow }^{\dagger }\right) ,  \notag \\
N &=&\left( \frac{1-|z_{1}|^{4}}{2}\right) ^{1/2}.  \label{gam}
\end{eqnarray}
Similarly, interchanging $\uparrow $ and $\downarrow $ and inverting the
signs of $\lambda $ and $\Delta $ we have

\begin{equation}
\gamma _{\downarrow }=N\sum_{n=0}^{\infty }\bar{z}_{1}^{2n}\left(
c_{2n+1\downarrow }+pc_{2n+1\uparrow }^{\dagger }\right) , \label{gamd}
\end{equation}%
which corresponds to the Kramers partner of Eq. (\ref{gam}).
Surprisingly, the even sites do not enter these Eqs.   In addition, 
note that

\begin{equation}
\gamma _{\uparrow }^{\dagger }=\bar{p}\gamma _{\downarrow }\text{, }\;\;\;\;\;\; \gamma
_{\downarrow }^{\dagger }=\bar{p}\gamma _{\uparrow }\text{.}  \label{conj}
\end{equation}%
Then, it is possible to define two independent Majorana operators

\begin{equation}
\eta_{1}=\gamma _{\uparrow }^{\dagger }+\gamma _{\uparrow }\text{, } \;\;\;\;\;\;
\eta_{2}=i(\gamma _{\uparrow }^{\dagger }-\gamma _{\uparrow }),  \label{majo}
\end{equation}%
such that $\eta_{i}^{\dagger }=\eta_{i}$.

A particular simple case is $t\rightarrow 0$, $|\Delta |=|\lambda |$, for
which $z\rightarrow 0$ [see Eq.  (\ref{z2})] and then Eqs.  (\ref{gam}), 
(\ref{gamd}) reduce to

\begin{equation}
\gamma _{\sigma }=\frac{1}{\sqrt{2}}\left( c_{1\sigma }+pc_{1-\sigma
}^{\dagger }\right) .  \label{simpl}
\end{equation}%
It can be verified directly that for these parameters, 
$[\gamma _{\sigma},H]=0$.

We can proceed in a similar way to derive the zero-energy excitations localized at the right end. 
Using Eqs.  (\ref{eigb}), (\ref{vsu}), and (\ref{zs}) we obtain

\begin{equation}
\frac{v_{3}}{u_{3}}=\frac{v_{4}}{u_{4}}=-p.  \label{vsu3}
\end{equation}%
Then, using Eqs.  (\ref{ep2}), the right-end eigenstates of zero energy can be expressed  in terms of fermionic 
operators as 

\begin{eqnarray}
\tilde{\gamma}_{\uparrow } &=&N\sum_{n=0}^{\infty }\bar{z}_{1}^{2n}\left(
c_{L-2n\uparrow }-pc_{L-2n\downarrow }^{\dagger }\right) ,  \notag \\
\tilde{\gamma}_{\downarrow } &=&N\sum_{n=0}^{\infty }z_{1}^{2n}\left(
c_{L-2n\downarrow }-pc_{L-2n\uparrow }^{\dagger }\right) .  \label{gamt}
\end{eqnarray}
Similarly to the left-end excitations, 
they are  related by

\begin{equation}
\tilde{\gamma}_{\uparrow }^{\dagger }=-\bar{p}\tilde{\gamma}_{\downarrow }%
\text{, }\tilde{\gamma}_{\downarrow }^{\dagger }=-\bar{p}\tilde{\gamma}%
_{\uparrow }.  \label{conj2}
\end{equation}

\subsection{Properties of the end excitations of the infinite chain}
\label{prop}

The relations given by Eqs. (\ref{conj}) and  (\ref{conj2}) imply
\begin{equation}
\{\gamma_{\uparrow },\gamma_{\downarrow }\}=p, \,
\{\tilde{\gamma}_{\uparrow },\tilde{\gamma}_{\downarrow}\}=-p.
\label{conmu}
\end{equation}

Denoting by $\hat{\gamma}_\sigma$ any of the two operators $\gamma_\sigma$, $\tilde{\gamma}_\sigma$, it is easy to 
verify that
\begin{equation}
[\hat{\gamma}_\sigma,S_z]=(s/2)\hat{\gamma}_\sigma,
\label{conmuz}
\end{equation}
where $s=1$ (-1) for $\sigma=\uparrow$ ($\downarrow$) and 
$S_z= \sum_i (c_{i \uparrow}^\dagger c_{i \uparrow} - c_{i \downarrow}^\dagger c_{i \downarrow})/2$ is 
the total spin projection in the Rashba direction $z$. Using Eq. (\ref{conmuz}) it is easy to see that
rotating the operator $\hat{\gamma}_\sigma$ an angle $\varphi$ around $z$ one obtains
\begin{equation}
\exp{(-i \varphi S_z)}\hat{\gamma}_\sigma\exp{(i \varphi S_z)}=\exp{(s i \varphi/2)\hat{\gamma}_\sigma}.
\label{rotz}
\end{equation}
Then $\hat{\gamma}_\uparrow$ ($\hat{\gamma}_\downarrow$) transforms like a spin with $S_z=-1/2$ (1/2) 
under rotations around $z$. 
 Note however that due to the Rashba spin-orbit coupling, the total spin is not conserved. For example one has
\begin{equation}
[\gamma_\uparrow,S_x]=\frac{N}{2} 
\sum_{n=0}^{\infty }z_{1}^{2n}\left( c_{2n+1\downarrow}-pc_{2n+1\uparrow }^{\dagger }\right),
\label{conmux}
\end{equation}
and the second member anticommutes with all low-energy operators  
$\hat{\gamma}_\sigma$. This is also true for $S_y$ and the rest of the $\hat{\gamma}_\sigma$ operators. 

The end excitations described by $\gamma_{\sigma}, \tilde{\gamma}_{\sigma}$ satisfy the usual Pauli exclusion principle
$\gamma_{\sigma}^2=\tilde{\gamma}_{\sigma}^2=0$.  Hence, taking into account the relations of  Eqs. (\ref{conj}) and  (\ref{conj2}), we can characterize
the end states by two operators, $\gamma \equiv \gamma_{\uparrow}$ and $\tilde{\gamma} \equiv \tilde{\gamma}_{\uparrow}$. These operators define
q-bit states  $|L 0\rangle$, $|L1\rangle$, localized at the left edge and $|R 0\rangle$, $|R1\rangle$, localized at the right, satisfying
\begin{eqnarray} \label{qbits}
&  \gamma|L 0\rangle=0, \;\;\;\;\;\;\;\; |L1\rangle=\gamma^{\dagger}|L 0\rangle, \nonumber \\
& \tilde{\gamma}|R 0\rangle=0,\;\;\;\;\;\;\;\; |R1\rangle=\tilde{\gamma}^{\dagger}|R 0\rangle.
\end{eqnarray}

Eqs. (\ref{z2}), (\ref{gam}), (\ref{gamd}) and (\ref{gamt}) define the zero-energy excitations
of the infinite chain.
For finite odd $L$, the corresponding operators continue to commute with the Hamiltonian.\cite{note}
This means that the zero energy excitations persist. This is a particular 
property of the case $\mu=0$. For even finite $L$, the states $\gamma_{\sigma}$
and $\tilde{\gamma}_{\sigma}$ mix as described in the next section.

\subsection{Extension to finite large even $L$}
\label{lfin}

For finite odd $L$, as long as $\mu=0$, the edge states have the same properties as those of $L \rightarrow \infty$. Namely, they have exactly energy $E=0$, and they can be expressed as in 
Eqs.  (\ref{gam}), (\ref{gamd}) and (\ref{gamt}). 
For finite even $L$, the left and right zero modes discussed in the previous section
hybridize and the resulting subgap eigenstates have  finite energy $\pm E$,
which decreases exponentially with $L$. For even $L$, the first correction
to the solution presented in the previous section is of order $E\sim |z_{1}|^{L}$. 
From Eq.  (\ref{e2})
we see that the first correction to the roots $z_{i}$ is of order $E^{2}$. 
Then, to linear order in $E$, the $z_{i}$ are not modified. On the other hand Eqs.  (\ref{eigb}), (\ref{vsu}) are
 linear in $E$. Explicitly, after substituting the solution of Eq. (\ref{z2}) and the definition of Eq. (\ref{pe})
 they read

\begin{eqnarray}
\frac{v_{1}}{u_{1}} &=&p+\frac{e^{i\phi }E}{\Delta \left( z_{1}+\frac{1}{%
z_{1}}\right) },  \notag \\
\frac{v_{2}}{u_{2}} &=&p-\frac{e^{i\phi }E}{\Delta \left( z_{1}+\frac{1}{%
z_{1}}\right) },  \notag \\
\frac{v_{3}}{u_{3}} &=&-p+\frac{e^{i\phi }E}{\Delta \left( \bar{z}_{1}+%
\frac{1}{\bar{z}_{1}}\right) },  \notag \\
\frac{v_{4}}{u_{4}} &=&-p-\frac{e^{i\phi }E}{\Delta \left( \bar{z}_{1}+%
\frac{1}{\bar{z}_{1}}\right) }.  \label{vsue}
\end{eqnarray}%
As we know from the previous Section, for $L\rightarrow \infty $, $E=0$, the
two ends are decoupled and Eqs.  (\ref{ep1}), (\ref{ep2}) give $\alpha
_{2}=\alpha _{1}$, $\alpha _{4}=\alpha _{3}$ for the coefficients of the
eigenstate $\sum \alpha _{k}|e_{k}\rangle $. We define two deviations from this limit,
linear in $E$, $\beta _{1}=\alpha _{2}-\alpha _{1}$, $\beta _{3}=\alpha
_{4}-\alpha _{3}$. Choosing $w_{k}=1/u_{k}$ for $k=1,2$ (weight 1 for 
$c_{1\uparrow }$) and $w_{k}=z_{k}^{1-L}/u_{k}$ for $k=3,4$ (weight 1 for 
$c_{L\uparrow }$), the linear corrections in $E$ to Eqs.  (\ref{ep1}) lead to the following equations \cite{note}

\begin{eqnarray}
0 &=&2\alpha _{1}E\left[ \frac{z_{1}}{\left( z_{1}+\frac{1}{z_{1}}\right) }-1%
\right]  \notag \\
&&+2\alpha _{3}\bar{z}_{1}^{L-2}\left[ -t+i\lambda -pe^{-i\phi }\Delta %
\right]  \notag \\
&&+\beta _{1}z_{1}\left[ t-i\lambda -pe^{-i\phi }\Delta \right] ,
\label{c1}
\end{eqnarray}

\begin{eqnarray}
0 &=&-2\alpha _{1}E\left[ p+\frac{\left( -t+i\lambda \right) e^{i\phi
}z_{1}}{\Delta \left( z_{1}+\frac{1}{z_{1}}\right) } \right]  \notag \\
&&+2\alpha _{3}\bar{z}_{1}^{L-2}\left[ \left( -t+i\lambda \right)
p+e^{i\phi }\Delta \right]  \notag \\
&&-\beta _{1}z_{1}\left[ p\left( t-i\lambda \right) +e^{i\phi }\Delta \right].  
\label{c2}
\end{eqnarray}

From these expressions, $\beta_1$ can be eliminated leaving an equation that relates $\alpha_1$ and $\alpha_3$.
In fact $\beta_1$ turns out to be exponentially small and we neglect it.
More precisely, performing the operation Eq. (\ref{c2}) + Eq. (\ref{c1})$/p$ 
and substituting  Eq.  (\ref{pe}),
results in the following relation between $\alpha _{1}$ and $\alpha _{3}$:

\begin{equation}
Ey\alpha _{1}+x\alpha _{3}=0,  \label{ece1}
\end{equation}
where
\begin{eqnarray}
y &=&1+\frac{i\text{sign}(\Delta \lambda )(t-i\lambda )z_{1}^{2}+\Delta }{%
\left( z_{1}^{2}+1\right) \Delta },  \notag \\
x &=&\bar{z}_{1}^{L-2} \varepsilon, \;\;\;\;\;\;\;\;\;\; \varepsilon= 2 \left[ t-i\lambda +i\text{sign}(\Delta \lambda
)\Delta \right] .  \label{yx}
\end{eqnarray}%

We can follow a similar procedure to evaluate  the linear corrections in $E$ to Eqs.  (\ref{ep2}). The result is

\begin{equation}
\bar{x}\alpha _{1}+E\bar{y}\alpha _{3}=0.  \label{ece2}
\end{equation}%
From Eqs.  (\ref{ece1}) and (\ref{ece2}) we finally obtain the desired energy and the ratio $\alpha _{3}/\alpha _{1}$ which has modulo 1. 
For the energy, we get 

\begin{equation}
E=\pm \left\vert \frac{x}{y}\right\vert  = \pm \left\vert \frac{\bar{z}_{1}^{L-2} \varepsilon}{y}\right\vert ,  \label{ene}
\end{equation}%
which explicitly defines a relation between the finite length of the chain and the non-zero energy of the excitations. For positive $E$ we define $\theta $ from

\begin{equation}
\frac{\alpha _{3}}{\alpha _{1}}=-\frac{\bar{x}}{\bar{y}E}=e^{i\theta }.
\label{theta}
\end{equation}%

We recall that $\alpha_1\; (\alpha_3)$ is the amplitude of the quasiparticle excitation at the left (right) end of the chain. The corresponding excitations are described 
by $\gamma_{\sigma}, \; \tilde{\gamma}_{\sigma}$ given by Eqs.  (\ref{gam}) and (\ref{gamt}). In the case of the finite chain we are analyzing here, the exact subgap 
eigenstate with energy $E >0$ given by Eq. (\ref{ene}),  is a linear combination of the latter ones. 
It can be described in terms of 
 the  annihilation operator of a quasiparticle with this energy as 
\begin{equation}
\Gamma _{\uparrow }=\frac{1}{\sqrt{2}}\left( \gamma _{\uparrow }+e^{i\theta }%
\tilde{\gamma}_{\uparrow }\right) ,  \label{gamcu}
\end{equation}%
where $\gamma _{\uparrow }$ and $\tilde{\gamma}_{\uparrow }$ have the same
form as in Eqs. (\ref{gam}) and (\ref{gamt}), except for the fact that for the finite chain the
sum in these equations extends up to $L$ instead of $\infty$, and then the
normalization changes to
\begin{equation}
N=\left[ \frac{1-|z_{1}|^{4}}{2\left( 1-|z_{1}|^{2L}\right) }\right] ^{1/2}.
\label{n2}
\end{equation}%
As before, a degenerate solution is obtained interchanging spin up and down
and changing the sign of both $\Delta $ and $\lambda $ (or by time reversal operation
if $\phi =0$)

\begin{equation}
\Gamma _{\downarrow }=\frac{1}{\sqrt{2}}\left( \gamma _{\downarrow}+e^{-i\theta }\tilde{\gamma}_{\downarrow }\right) .  \label{gamcd}
\end{equation}

Using Eqs.  (\ref{conj}) and (\ref{conj2}), it can be easily checked that the
low-energy eigenstates with negative energy given by Eq. (\ref{ene}) 
coincide except for an irrelevant factor with the operators 
$\Gamma_{\uparrow }^{\dagger }$ and $\Gamma _{\downarrow }^{\dagger }$, transpose conjugate 
of those defined by Eqs. (\ref{gamcu}) and Eqs. (\ref{gamcd}). This can be expected since  
taking the transpose conjugate of the equation 

\begin{equation}
\lbrack \Gamma _{\sigma },H \rbrack =E\Gamma _{\sigma },  
\label{eigen}
\end{equation}
implies $\lbrack \Gamma _{\sigma }^{\dagger },H \rbrack = -E\Gamma _{\sigma }^{\dagger }$.

Here and in what follows, when an operator $O$ satisfies $\lbrack O,H \rbrack =E O$, 
with $E \neq 0$, implying $\lbrack O^{\dagger },H \rbrack = -E O^{\dagger }$ 
we choose as annihilation operator $O$ for positive $E$ and $O^{\dagger }$ for negative $E$, so that the vacuum
$|0\rangle_O $ of all these annihilation operators ($O |0\rangle_O =0$) is the ground state.

\section{Localization length of the end states}

\label{loca}

\begin{figure}[t]
\begin{center}
\includegraphics[width=\columnwidth]{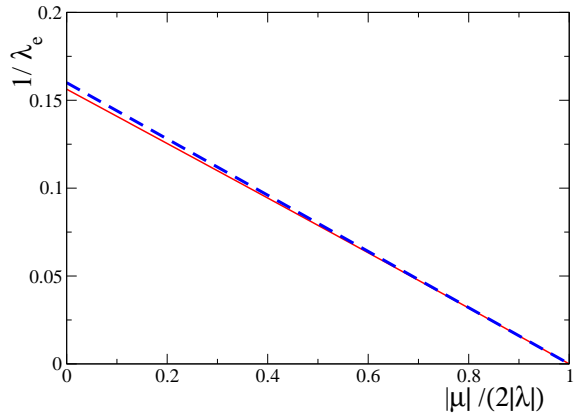}
\end{center}
\caption{(Color online) Full line: Inverse of the localization length as a
function of chemical potential. Dashed line corresponds to Eq, (\ref{longe}). 
Parameters are $t=1$, $\Delta =0.5$, $\lambda =0.2$.}
\label{long}
\end{figure}

\begin{figure}[t]
\begin{center}
\includegraphics[width=\columnwidth]{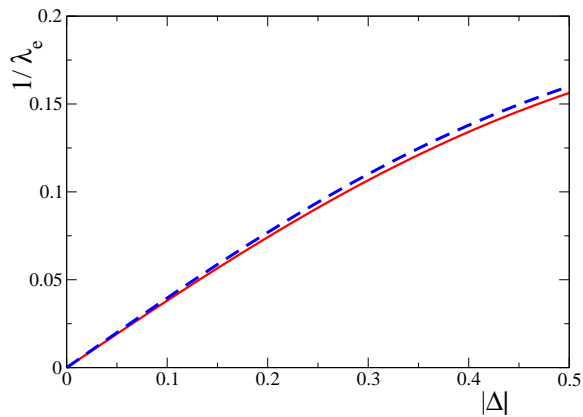}
\end{center}
\caption{(Color online) Full line: Inverse of the localization length as a
function of $\Delta$. Dashed line corresponds to Eq, (\ref{longe}). 
Parameters are $t=1$, $\mu =0$, and $\lambda =0.2$.}
\label{long2}
\end{figure}

To define the localization length $\lambda _{e}$ of the end states, it
suffices to consider a chain of infinite length. In this case the energy of
the low energy states is $E=0$ and it is not necessary to solve the boundary equation to obtain
$E$.
The four complex roots of Eq. (\ref{e2})
provide the decay along the chain of the components of the eigenstates of the
Hamiltonian, as illustrated in Sec. \ref{alasec}. We choose the eigenstates
localized at the left end for the following discussion (of course the
results are the same choosing the right end). From the four solutions of Eq. (\ref{e2}), 
only those two with $|z_{i}|<1$ contribute to the states localized at the left end. Let
us denote as $|z_{l}|$ the largest of these two absolute values. Clearly
this is the one that determines the localization length because at large
distances, the probability $p(n)$ of finding a particle at site $n$ is
proportional to $|z_{l}|^{2n}$. Defining as usual the localization length 
$\lambda _{e}$, from $p(n)\sim \exp (-n/\lambda _{e})$ we obtain

\begin{equation}
\lambda _{e}=\frac{-1}{2\ln |z_{l}|}.  \label{le}
\end{equation}

For $\mu=0$ both roots with $|z_{i}|<1$ are given by Eq. (\ref{z2}) and replacing 
in Eq. (\ref{le}), $\lambda _{e}$ is derived. 
In general, one has to solve the quartic equation (\ref{e2}) and choose the largest $|z_{i}|$ 
with the condition $|z_{i}|<1$ to obtain the localization length.
Following this procedure we derived the results shown in Fig. \ref{long}, where the localization length as a 
function of the chemical potential $\mu$ is represented. 
Note that if $z$ is a solution of
Eq. (\ref{e2}), $-z$ is a solution for the opposite value of $\mu $. This
and other properties listed above Eq. (\ref{e2}) allows us to restrict the
calculations and discussions to all parameters assumed positive, and extend later 
the result for all signs using the 
symmetry properties of Eqs. (\ref{e2}) and (\ref{le}). Starting from $\mu =0$ and increasing 
$\mu $, the localization length increases and diverges as the transition
to the non-topological phase at $\mu =2\lambda $ is approached, as expected.
In fact for $\mu =2\lambda $ there is a double root $z=-i$ of Eq. (\ref{e2})
for $E=0$. The other two roots are given by

\begin{equation}
z=\frac{i\left( t^{2}+\Delta ^{2}+\lambda ^{2}\pm 2\Delta \lambda \right) }
{t^{2}+\Delta ^{2}-\lambda ^{2}-2it\lambda }.  \label{zother}
\end{equation}
It is easy to see that for positive $\Delta$ and $\lambda$, one of these two roots has
$|z|>1$ and the other $|z|<1$. 
Then for $\mu \rightarrow 2\lambda $, $|z_{i}|\rightarrow 1$, and $\lambda
_{e}\rightarrow +\infty $.

In order to find an analytical expression for the localization length near the topological
transition, we expand Eq. (\ref{e2}) up to total second order in $y=z+i$ and 
$\epsilon =2\lambda -\mu >0$ around the point $y=\epsilon =0$, to obtain after some
algebra

\begin{equation}
4(t^{2}+\Delta ^{2})y^{2} -4 t y \epsilon +\epsilon ^{2}=0.
\label{e2a}
\end{equation}
Solving this equation, using Eq. (\ref{le}) and extending the results to
other signs of $\mu $ and $\lambda $ we get 
\begin{equation}
\lambda _{e}\simeq \frac{t^{2}+\Delta ^{2}}{|\Delta|(2|\lambda |-|\mu |)}.
\label{longe}
\end{equation}
While this equation is expected to be valid only near the topological
transition, it is a good approximation for the whole range of chemical
potential within the topological phase defined by $2|\lambda|>|\mu|$ and $\Delta \neq 0$, 
as indicated in Figs. \ref{long} and \ref{long2}. 
Note that $\lambda _{e}$ diverges not only at the boundary $|\mu| \rightarrow 2|\lambda|$ of the topological phase, 
but also for $\Delta \rightarrow 0$, which is of course also a boundary of the 
topological phase and in addition to superconductivity. 

For a chain of finite length $L$, on general physical grounds one expects that  the energy $E$ 
of the low-energy  excitations is proportional to $\exp (-L/\lambda _{e})$.
This is supported by the analytical results of the previous section for $\mu=0$,

\section{Spin distribution  along a finite chain}
\label{spin}

In this Section, we calculate the density of spin projection along the Rashba direction $z$  
at each site of a finite long chain.
We show that for subgap excitation with an odd number of particles, the ground state has a localized
spin projection $S_e^z= \pm 1/4$ at each end.

For an even number of particles, the ground state $|g_{e}\rangle $ can be
constructed  by applying to the vacuum $|0\rangle_c $ of the $c_{i\sigma }$ operators
($c_{j\sigma }|0 \rangle_c =0$)
a product of all annihilation operators of excitations that satisfy 

\begin{equation}
\lbrack \Gamma _{\nu },H]=E_{\nu }\Gamma _{\nu },  \label{gnu}
\end{equation}%
with positive $E_{\nu }$. Denoting $\Gamma _{\Theta \nu }=\Theta \Gamma _{\nu
}\Theta^{-1 }$, where $\Theta$ is the time reversal operator, the ground state
can be written in the form

\begin{equation}
|g_{e}\rangle =\tilde{N}\prod\limits_{\nu }\Gamma _{\nu }\Gamma _{\Theta \nu
}|0\rangle_c ,  \label{ge}
\end{equation}%
where $\tilde{N}$ is a normalization factor.

The $z$ component of the spin operator at site $j$ is

\begin{equation}
S_{j}^{z}=\frac{c_{j\uparrow }^{\dagger }c_{j\uparrow }-c_{j\downarrow
}^{\dagger }c_{j\downarrow }}{2}.  \label{sz}
\end{equation}%
Therefore

\begin{eqnarray}
\langle S_{i}^{z}\rangle  &=&\langle g_{e}|S_{i}^{z}|g_{e}\rangle =\langle
g_{e}|\Theta^{-1 } \Theta S_{i}^{z} \Theta^{-1 } \Theta |g_{e}\rangle   \notag \\
&=&-\langle g_{e}|S_{i}^{z}|g_{e}\rangle =0,  \label{kk}
\end{eqnarray}
hence, the expectation value of the spin projection at each site 
vanishes in the ground state of  a chain with an even
number of particles.

The ground state for an odd number of particles, corresponds to creating the one-particle excitation
of lowest energy to $|g_e\rangle$. The energy cost is rather large for all $\Gamma _\nu^\dagger$ except for 
the subgap states close to $E=0$ calculated
in the previous Section (or their generalization for finite $\mu $). The latter have
energy $E\sim |z_{1}|^{L-2}$ decaying exponentially with $L$. In order to split
the spin degeneracy we assume that a small magnetic field is applied, so
that the ground state becomes

\begin{equation}
|g_{o}\rangle =\Gamma _{\uparrow }^{\dagger }|g_{e}\rangle,  \label{go}
\end{equation}
where 
\begin{equation}
\Gamma _{\uparrow }=\sum\limits_{j}\left( u _{j  }c_{j\uparrow }+v_{j }c_{j\downarrow }^{\dagger }\right) ,  \label{gams}
\end{equation}
with the coefficients  given by Eqs.  (\ref{gam}), (\ref{gamt}) and (\ref{gamcu}) for $\mu =0$.
Then

\begin{eqnarray}
\langle S_{i}^{z}\rangle  &=&\langle g_{e}|\Gamma _{\uparrow
}S_{i}^{z}\Gamma _{\uparrow }^{\dagger }|g_{e}\rangle   \notag \\
&=&\langle g_{e}|[\Gamma _{\uparrow },S_{i}^{z}]\Gamma _{\uparrow }^{\dagger
}|g_{e}\rangle ,  \label{so1}
\end{eqnarray}%
where we have used Eq.  (\ref{kk}) in the last equality.

From Eqs. (\ref{sz}) and (\ref{gams}) we get

\begin{equation}
\lbrack \Gamma _{\uparrow },S_{i}^{z}]=\frac{u _{i }c_{i\uparrow }+v_{i }c_{i\downarrow }^{\dagger }}{2}  \label{conm}
\end{equation}

Using $\langle g_{e}|\Gamma _{\uparrow }^{\dagger }=0$ we can write Eq.  (\ref%
{so1}) in the form

\begin{equation}
\langle S_{i}^{z}\rangle =\langle g_{e}|\{[\Gamma _{\uparrow
},S_{i}^{z}],\Gamma _{\uparrow }^{\dagger }\}|g_{e}\rangle ,  \label{antic}
\end{equation}%
and calculating the anticommutators using Eqs.  (\ref{gams}) and (\ref{conm})
we finally obtain

\begin{equation}
\langle S_{i}^{z}\rangle =\frac{|u _{i }|^{2}+|v _{i }|^{2}}{2}
\label{sof}
\end{equation}

\subsection{Examples}
\begin{figure}[h]
\begin{center}
\includegraphics[width=\columnwidth]{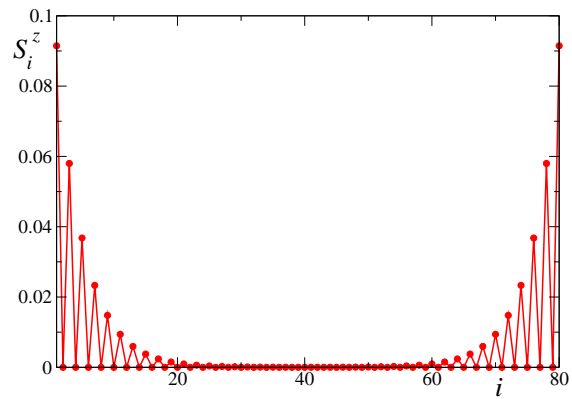}
\end{center}
\caption{Spin projection at each site for an odd number of particles and total spin projection 1/2.
Parameters are $\mu=0$, $t=1$, $\Delta=0.5 $, $\lambda=0.2$ and $L=80$.}
\label{mu0}
\end{figure}

In Fig. \ref{mu0} we show  $\langle S_{i}^{z}\rangle$ for an odd number of particles and total spin 
projection $S_z=\sum_i S_{i}^{z}=1/2$, obtained from the analytical solution, as described in the previous Section.
We also checked the result by numerical diagonalization of the finite chain. The results are practically identical.
The energy of the excitation (which coincides with the difference in energy of the ground state for odd an even 
particles) is $E=6.314 \times 10^{-4}$. The numerical energy is higher by $\sim 2 \times 10^{-8}$ which can be 
ascribed to terms of order $E^2$ neglected in the analytical treatment.

\begin{figure}[h]
\vspace{1.cm}
\begin{center}
\includegraphics[width=\columnwidth]{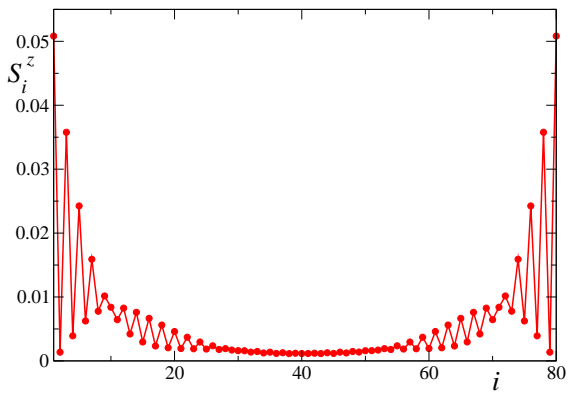}
\end{center}
\caption{Same as Fig. \ref{mu0} for $\mu=0.2$.}
\label{mup2c}
\end{figure}

\begin{figure}[t]
\begin{center}
\includegraphics[width=\columnwidth]{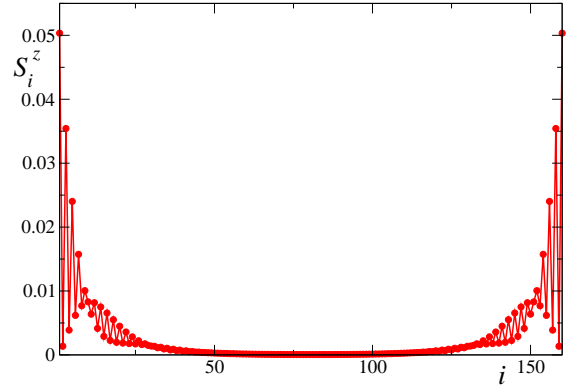}
\end{center}
\caption{Same as Fig. \ref{mup2c} for $L=160$.}
\label{mup2l}
\end{figure}

Clearly half of the total spin projection is localized at each end of the chain 
and $\langle S_{i}^{z}\rangle$ is practically zero in the middle of the chain. 
Although the physics is different, this is reminiscent of the spin 1/2 excitations at
the ends of the $S=1$ antiferromagnetic chain.\cite{miya,white,bati} In addition, there is a marked even-odd 
oscillation. While $\langle S_{i}^{z}\rangle$ decays exponentially as the distance from the ends increase, 
$\langle S_{i}^{z}\rangle$  vanishes exactly
at distances equal to an odd number of lattice constants from the ends. 
This is a particular property of the case $\mu=0$, but the oscillations remain for 
finite $\mu$ as shown in Figs. \ref{mup2c} and \ref{mup2l}.

In Fig. \ref{mup2c} we show $\langle S_{i}^{z}\rangle$ as a function of the site, derived form the numerical
solution of the chain for the same parameters as before, except for the fact that $\mu$ 
is increased but 
not too much, in order that the system is kept within the topological region $|\mu |< 2 |\lambda| $.
In this case, according to the calculations of the previous section, the localization length 
of $\langle S_{i}^{z}\rangle$ at the ends of the chain increases from  $\lambda_e \simeq 6.4$ to 12.7.  
This is consistent with the increase in the energy of the 
excitation by nearly an order of magnitude to $E=5.51 \times 10^{-3}$. In contrast to the case $\mu=0$ for which 
the energy vanishes for chains of odd length, the energy is similar for one site less ($E=5.71 \times 10^{-3}$) or one 
site more ($E=5.28 \times 10^{-3}$). Actually, for an homogeneous chain, the period of the oscillations is given by the Fermi wavelength of the system without superconductivity, which in turn depends
on the chemical potential. 
For the chosen length of the chain $L=80$, with order of magnitude comparable to $\lambda_e \simeq 12.7$, 
the spin excitations at the ends are not well separated.  
However, the overall trend is similar to the previous
case, with larger $\langle S_{i}^{z}\rangle$ at the ends and even-odd oscillations. 
Curiously, while for $i \leq 10$, $\langle S_{i}^{z}\rangle$
is larger for odd sites, the situation is reversed for $11  \leq i \leq 30$. A similar situation takes place at the 
other end replacing $i$ by $L+1-i$.

As shown in Fig. \ref{mup2l} if the length of the chain is increased while keeping the same energy parameters, 
$\langle S_{i}^{z}\rangle$ near the ends is practically not affected. However it is now clear that the 
spin excitations at both ends are well separated. In this case, the excitation energy is $E=2.335 \times 10^{-4}$.

\section{General properties of the end states}
While the specific pattern of the spacial distribution of the spin density or the explicit expression for the 
localization length depends on the model parameters, as discussed in the previous section, there are
other features of the topological phase, which are much more general and only depend on the symmetries of the model. 
In this section, we focus on such features and we summarize the main properties of the subgap excitations 
that are valid for any 1D TRITOPS system conserving a given spin porjection, in our case $S_z$.
In particular  Eqs. (\ref{conmu}) which we reproduce here for the ease of the reader
\begin{equation}
\{\gamma_{\uparrow },\gamma_{\downarrow }\}=p, \,
\{\tilde{\gamma}_{\uparrow },\tilde{\gamma}_{\downarrow}\}=-p,
\end{equation}
where $p$ is defined in Eq. (\ref{pe}) have been demonstrated using a continuum formulation in Section II  of the supplemental material of 
Ref. \onlinecite{cam}) for $\phi=0$. We expect them to be generally valid. Extension for $\phi \neq 0$ is trivial using
a gauge transformation.

These operators obey the commutation rules
with the operator $S_z$ (with $z$ being the direction of the spin-orbit interaction) that were  given in 
Eq. (\ref{conmuz}).

In the case of a finite chain with length $L$, 
the exact subgap 
eigenstate with energy $E >0$ given by Eq. (\ref{ene}),  is a linear combination of the form given in Eq. (\ref{gamcu}),
\begin{equation}
\Gamma _{\uparrow }=\frac{1}{\sqrt{2}}\left( \gamma _{\uparrow }+e^{i\theta }%
\tilde{\gamma}_{\uparrow }\right),
\end{equation}%
which satisfies  $\{\Gamma_{\uparrow },\Gamma_{\downarrow }\}=0$.
The time-reversal partner is the operator defined in Eq. (\ref{gamcd}),
\begin{equation}
\Gamma _{\downarrow }=\frac{1}{\sqrt{2}}\left( \gamma _{\downarrow}+e^{-i\theta }\tilde{\gamma}_{\downarrow }\right), 
\end{equation}
where $\theta$ depends on the model.
Generally, going from the infinite chain to the finite one introduces a tunneling between the operators
of both ends which can be written as $Ee^{i\theta }$ with $E$ real. This point will be further discussed in the 
next Section [see Eqs. (\ref{eqb})], whose results are also generic.
For the particular model of Section II, $\theta$ is given by Eq. (\ref{theta}). 

The operators $\Gamma_{\sigma}$ 
annihilate two degenerate excitations (corresponding to $\sigma= \uparrow, \downarrow$) with energy $E$. 
Let us also highlight that the structure of the operators $\Gamma_{\sigma}$ defined in
Eqs. (\ref{gamcu}) and (\ref{gamcd})  imply entanglement of the excitations localized at the left and the right ends of the chain. In fact, we can define  2 q-bit states in the basis
of Eq. (\ref{qbits}) $|l,r\rangle \equiv |lL, rR \rangle$, with $l,r=0,1$ and analyze the effect of generating 
two quasiparticles with $\Gamma_{\sigma}^{\dagger}$ on these states. The result is
\begin{eqnarray}
 \Gamma_{\uparrow}^{\dagger} \Gamma_{\downarrow}^{\dagger} |0,0 \rangle &=&\Gamma_{\uparrow}^{\dagger} 
 \Gamma_{\downarrow}^{\dagger} |1,1\rangle = 0, \nonumber \\
 \Gamma_{\uparrow}^{\dagger} \Gamma_{\downarrow}^{\dagger} 
 \frac{1}{\sqrt{2}} \left[ |0,1 \rangle \pm  |1,0  \rangle\right]  & = &
\frac{\overline{p}}{2\sqrt{2}} \left[ \left(\pm 1 - e^{i \theta}\right) |1,0 \rangle \right. \nonumber \\
& & \left. - \left(1 \mp e^{-i \theta}\right) |0,1  \rangle \right]. \label{bell}
\end{eqnarray}
We see that  $ \Gamma_{\uparrow}^{\dagger} \Gamma_{\downarrow}^{\dagger}$  does not act on the subspace of product states of single q-bits. On the other hand, since the phase $\theta$ resulting from 
Eq. (\ref{theta}) is in general different from $0, \pi$, this operator maps Bell states into combinations of Bell states.  \cite{been} Notice that this construction relies on the fact that for finite wires, the energy of
the subgap states is finite. In contrast, in the limit of $L \rightarrow \infty$, the four two-q-bit states are exactly degenerate at $E=0$ and two-quasiparticle excitations can be constructed with any  linear
combination of these states. These operators, obey the commutation rules with the Hamiltonian given by Eq. (\ref{eigen}). 
Finally, it is interesting to notice that these properties are very similar to those discussed in the context of 
topological phases taking place in 1D spin systems. \cite{frac1,frac2}

\section{Effect of a magnetic field at one end}

\label{field}

\subsection{Quasiparticles}

In this Section we calculate the spin projection along the Rashba direction 
at the ends of the chain $S_e^z$ of
the ground state and low-energy excitations, as well as the energy
of these excitations under the effect of a weak magnetic field (or Ising type interaction) 
$B>0$ applied at one end only, parallel to the Rashba direction $z$.  
Without loss of generality we assume that the field is applied at the
right end of the chain and includes all sites for which $\langle
S_{i}^{z}\rangle $ is significantly different from 0 for $B=0$, which 
means that the length of the region subject to the magnetic field is much
larger than the localization length $\lambda_e$ of the low-energy
quasiparticles (see Sec. \ref{loca}). In particular it can include the right half of the chain. For concreteness,  
we assume the latter option
and write the Hamiltonian for the superconducting wire in the presence of the external magnetic field $B$, described by
a Zeeman coupling $\Delta_Z=g \mu_B B$ as
\begin{equation} \label{htot}
H_{\rm tot}= H-  \Delta_Z S^z_R,
\end{equation}
where $S^z_R= \sum_{j=L/2+1}^L S^z_j$ is the operator of the spin projection at the right half of the chain. 
A sketch is shown in Fig. \ref{fig0} (see bottom left). Similarly for the left end $S^z_L=S_z-S^z_R$.
Alternatively, we can also consider a magnetic island or a magnetic adatom with classical magnetic moment 
$M$  close to the right end of the chain coupled with the spin through an Ising interaction 
$H_B= - J S^z_R M$. This is equivalent to the Hamiltonian of Eq. (\ref{htot}) upon 
identifying $J M \equiv \Delta_Z $. In the case of the magnetic field 
we assume that its magnitude  is much smaller than the critical field of the superconductor. 
In the case of the magnetic moment, we assume a weak coupling $J$, such that it  interacts with the subgap 
edge-state excitation but it does not 
induce low-energy Yu-Shiba-Rusinov states \cite{YSR} inside the superconducting gap. 
This assumption implies that we can restrict the effect of the magnetic field to the low-energy 
in-gap states.

Taking for the moment our analytical solution for $\mu=0$ using Eqs. (\ref{gamcu}), 
(\ref{gamcd}), (\ref{conj}), and (\ref{conj2}) we obtain

\begin{eqnarray}
\left[ \gamma _{\sigma },H_{\rm tot} \right]  &=&Ee^{is\theta }\tilde{\gamma}_{\sigma },
\notag \\
\left[ \tilde{\gamma}_{\sigma },H_{\rm tot} \right]  &=&Ee^{-is\theta }\gamma _{\sigma }-s 
\frac{\Delta_Z}{2}\tilde{\gamma}_{\sigma },  
\label{eqb}
\end{eqnarray}
where $s=1 \; (-1)$ for spin $\uparrow (\downarrow )$.

For our model with $\mu =0$, the explicit values of $E$ and $e^{i\theta }$
are given by Eqs. (\ref{ene}) and (\ref{theta}). However, we want to stress
that the \emph{form} of Eqs. (\ref{eqb}) is generally valid for any TRITOPS
chain: the zero modes of the chain in the limit $L \rightarrow \infty$, 
$\gamma _{\sigma }$ at the left
and $\tilde{\gamma}_{\sigma }$ at the right become mixed and split in the
finite chain by an effective hopping $Ee^{i\theta }$  (whose detailed value
depends on the particular system) for spin up, and time reversal symmetry
implies a hopping  $Ee^{-i\theta }$ for spin down. In addition, 
it can be easily seen from our analytical solution for $\mu =0$ [see Eq. (\ref{conmux}) 
and the sentence below it], that
to linear order in $B$ the end states for $L \rightarrow \infty$ do not split if
the magnetic field is applied perpendicular to the direction of 
the Rashba field ($z$ in our notation). For general TRITOPS
models a splitting for all directions of the magnetic field is expected,
but with a strong anisotropy.\cite{dumi}

Using a Bogoliubov transformation, two annihilation operators can be
defined such that

\begin{equation}
\left[ \Gamma _{B,\sigma },H_{\rm tot} \right] =E_{\sigma }\Gamma _{B,\sigma },  \label{eta}
\end{equation}%
with $E_{\sigma }>0$. Specifically

\begin{eqnarray}
\Gamma _{B,\sigma } &=&\alpha _{\sigma }\gamma _{\sigma }+\beta _{\sigma
}e^{is\theta }\tilde{\gamma}_{\sigma },  \notag \\
E_{\sigma } &=&r-s\frac{\Delta_Z}{4},\label{etau}
\end{eqnarray}
being
\begin{eqnarray}\label{param}
r &=&\sqrt{\left( \Delta_Z/4\right) ^{2}+E^{2}},  \notag \\
\alpha _{\uparrow}^{2} &=& \beta _{\downarrow }^{2}=\frac{1}{2}+\frac{\Delta_Z}{4r},  \notag \\
\alpha _{\sigma}^{2} + \beta _{\sigma}^{2}  &=& 1, \, \alpha _{\sigma },\beta _{\sigma } > 0.  
\label{rab}
\end{eqnarray}

Using Eqs. (\ref{conmu}) and (\ref{rab}) one can verify that the operators  $\Gamma_{B,\sigma }$ 
and $\Gamma_{B,\sigma }^\dagger$ obey canonical anticommutation rules.

\subsection{Zeeman splitting}

The previous equations make explicit the fact that the finite energy $E$ of the excitations in chains of finite length 
has associated an hybridization of the localized zero modes. This implies a degree of entanglement between modes
localized at opposite ends. Our goal is to analyze the impact of this entanglement in the magnetic response.

The behavior of the quasiparticle excitations given by Eq. (\ref{etau}) have two important limits, 
which correspond to $\Delta_Z \gg E$  and $\Delta_Z \ll E$. 
\subsubsection{$\Delta_Z \gg E$}
This  corresponds to strongly localized end states with energy $E \sim 0$. This situation is achieved for very long chains, where the end modes
are almost completely decoupled.
 In this case we can expand $r$  defined in Eq. (\ref{param}) as 
$r \sim (\Delta_Z/4) \left[ 1 + (4 E/\Delta_Z)^2/2 \right] $ and we get 
\begin{eqnarray}
E_{\uparrow}&=& \frac{2 E^2}{\Delta_Z},\nonumber \\
E_{\downarrow}&=& \frac{2 E^2}{\Delta_Z}+ \frac{\Delta_Z}{2}.
\end{eqnarray}
In this limit, the operator $\Gamma _{B,\uparrow}$ that corresponds to the one-particle 
excitation with energy $E_{\uparrow}$ 
[see Eq. (\ref{etau})] tends  
to the quasiparticle $\gamma _{\uparrow}$  localized at the left 
end of the chain [$\alpha^2_\uparrow \sim 1$, see Eq. (\ref{rab})], 
and is not affected by the magnetic field.
In turn, the excitation with energy $E_{\downarrow}$, related to  
$\Gamma _{B,\downarrow} \sim \tilde{\gamma}_{\downarrow}$ 
with energy  $E_{\downarrow}$ corresponds 
to annihilating a quasiparticle at the right end of the chain with spin down [or creating one with spin up
since $\tilde{\gamma}_{\uparrow }^{\dagger }=-\bar{p}\tilde{\gamma}_{\downarrow }$, see Eq- (\ref{conj2})]. 
This leads to a decrease 
of the total energy in $E_{\downarrow} \sim \Delta_Z/2$, 
for annihilating an ordinary electron with spin down or creating one with spin up,
which is the expected result for an ordinary spin 1/2. 

Naturally, the complete spectrum of one-particle excitations also contains 
those corresponding to the Hermitian conjugate of the above described operators, in particular 
$\tilde{\gamma}_{\downarrow}^\dagger \sim \tilde{\gamma}_{\uparrow}$) with an energy 
\emph{loss} $E_{\downarrow}$, so that an ordinary Zeeman splitting is
$2E_{\downarrow} \sim \Delta_Z$
can be inferred from the magnetic-field dependence of the 
total spectral density of an ordinary electron observed in scanning tunneling spectroscopy, 
particularly if the STM tip 
is located near the end of the chain where the magnetic field is applied.

\subsubsection{$\Delta_Z \ll E$}
This case corresponds to a sizable hybridization and entanglement of the end modes. 
In this other limit we consider $r \sim E \left[ 1 + (\Delta_Z/4E)^2/2 \right] $. Hence
\begin{eqnarray}
E_{\uparrow}&=&  E \left[ 1 + \frac{1}{2}\left(\frac{\Delta_Z}{4E}\right)^2 \right]  - \frac{\Delta_Z}{4},\nonumber \\
E_{\downarrow}&=& E \left[ 1 + \frac{1}{2}\left(\frac{\Delta_Z}{4E}\right)^2 \right]  + \frac{\Delta_Z}{4}.
\end{eqnarray}
All low-energy quasiparticles have nearly equal weight at both ends 
[$\alpha^2_\uparrow \sim 1/2$, $\beta^2_\uparrow \sim 1/2$, see Eq. (\ref{rab})].
As a consequence, the effect magnetic field at only one end is reduced by a factor 1/2 with
respect to the application of the field in the whole sample.
The Zeeman splitting between the one-particle excitations of positive energy (corresponding to 
annihilation of quasiparticles) is  
$E_Z= E_{\downarrow}-E_{\uparrow}=\Delta_Z/2$, 
which is half the Zeeman splitting of a spin $1/2$.

We believe that this splitting might be observed 
not only by an STM which senses the one-particle spectral density but
also with microwave radiation 
which induces transtions conserving the number of electrons.\cite{tosi,hays}
While the light does not couple directly with the spin, the spin-orbit coupling couples it
with the orbital degrees of freedom and circularly polarized light induces transition
between stats with angular momentum projection 1/2 and -1/2.
As before, the full spectrum of one-particle excitations also contains negative energies 
with the same moduli as the positive ones described above.

\subsection{Spin polarization}

In our model for $\mu =0$ and large enough chains such that $|z_{1}|^{L}\ll 1$, 
using Eqs. (\ref{gamt}) and (\ref{n2}) one obtains that the low-energy
part of the spin projection at the right end $S_{R}^{z}=\sum_{i=L/2}^{L}S_{i}^{z}$ can be written in the form 

\begin{equation}
S_{R}^{z}\simeq \frac{1}{4}\left( \tilde{\gamma}_{\uparrow }^{\dagger }%
\tilde{\gamma}_{\uparrow }-\tilde{\gamma}_{\downarrow }^{\dagger }\tilde{%
\gamma}_{\downarrow }\right) ,  \label{szrb}
\end{equation}%
where we have neglected the contribution of the high-energy operators $%
\Gamma _{\xi },$ $\Gamma _{\xi }^{\dagger }$ with $[\Gamma _{\xi },H]=E_{\xi
}\Gamma _{\xi }$, where the subscript $\xi $ labels all operators with $E_{\xi }\gg E$. 
It is reasonable to expect that the low-energy part of $S_{R}^{z}$ has the same form for a general TRITOPS. 
Using Eqs. (\ref{etau}) and (\ref{conj2})
this part takes the following form, which is the most convenient one for
our purpose

\begin{equation}
S_{R}^{z}\simeq \frac{1}{4}\left( \alpha ^{2}-\beta ^{2}+2\beta ^{2}\Gamma
_{B\uparrow }^{\dagger }\Gamma _{B\uparrow }-2\alpha ^{2}\Gamma
_{B\downarrow }^{\dagger }\Gamma _{B\downarrow }\right) ,  \label{szrc}
\end{equation}%
where $\alpha =\alpha _{\uparrow }=\beta _{\downarrow }>0$, $\beta
=(1-\alpha ^{2})^{1/2}$ [see Eqs. (\ref{rab})].

As in section \ref{spin}, the ground state $|g_{e}\rangle $ for an even
number of particles is constructed by applying  to the vacuum of the 
$c_{j\sigma }$ all annihilation operators left invariant by the commutation
with the Hamiltonian:

\begin{equation}
|g_{e}\rangle =\tilde{N}\Gamma _{B\uparrow }\Gamma _{B\downarrow
}\prod\limits_{\xi }\Gamma _{\xi }|0\rangle _{c},  \label{geb}
\end{equation}
where  $\tilde{N}$ is a normalization factor.

Using Eqs. (\ref{szrc}) and (\ref{geb}) we get:

\begin{equation}
\langle g_{e}|S_{R}^{z}|g_{e}\rangle =\frac{\alpha ^{2}-\beta ^{2}}{4}=\frac{%
\Delta _{Z}}{4\sqrt{\left( \Delta _{Z}\right) ^{2}+16E^{2}}}.  \label{szre}
\end{equation}

For the states with odd number of particles $|g_{o\sigma }\rangle =\Gamma
_{B\sigma }^{\dagger }|g_{e}\rangle $, one obtains

\begin{equation}
\langle g_{o\sigma }|S_{R}^{z}|g_{o\sigma }\rangle =\frac{s}{4},  \label{szro}
\end{equation}
where $s=1\;(-1)$ for spin $\uparrow (\downarrow )$, independently of the
applied magnetic field at one end.
This fact is expected since for total $S_z$=1/2 or -1/2, there is only
one low-energy state and therefore it cannot be modified by a small
perturbation. The first correction is of order $(B/E_{\xi})^2$ and is
neglected in our approach.

\begin{figure}[h]
\begin{center}
\includegraphics[width=\columnwidth]{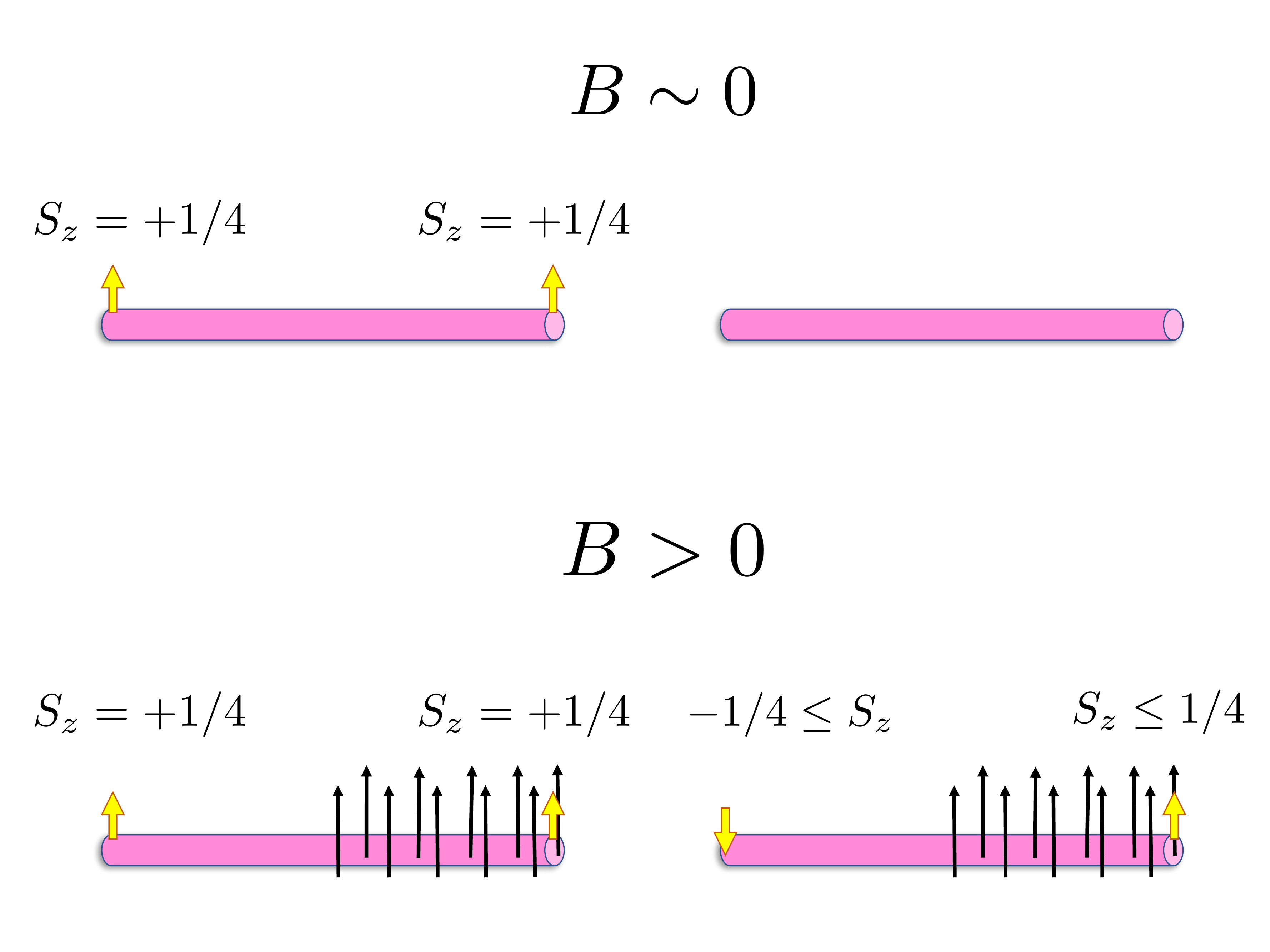}
\end{center}
\caption{Sketch of the ground-state expectation value of the spin at the ends 
when a magnetic field is applied at one of the ends of the wire. 
The left (right) panels of the figure correspond to 
a wire with an odd (even) number of particles.}
\label{figad}
\end{figure}

\begin{figure}[h]
\begin{center}
\includegraphics[width=\columnwidth]{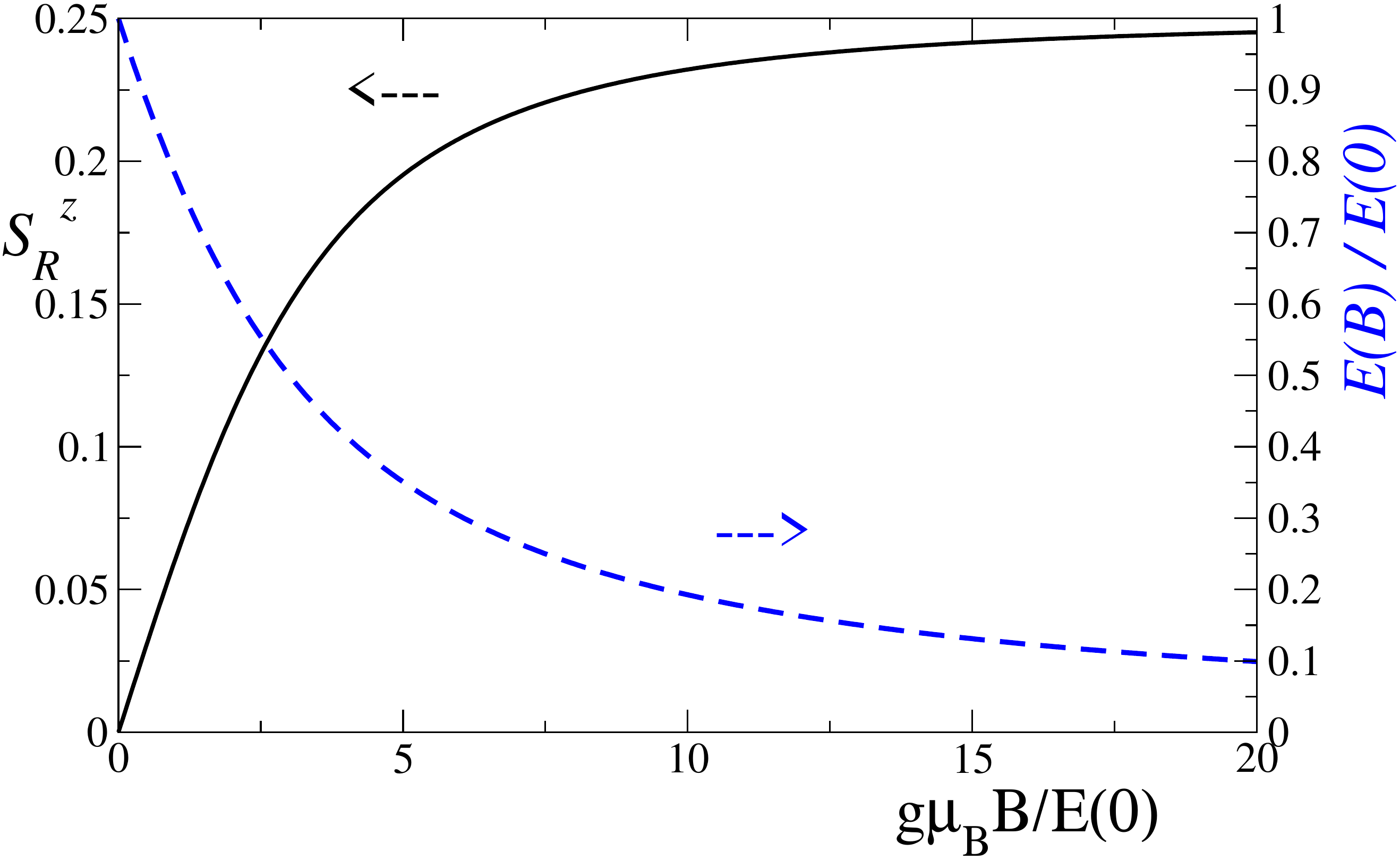}
\end{center}
\caption{(Color online) Ground-state spin projection at the right end of the chain (left scale full line) 
and first excitation energy (right scale dashed line) as a function of magnetic field applied to the right end.}
\label{spinb}
\end{figure}

In Fig. \ref{figad} we present sketches on the two different scenarios
expected in the magnetic response of wires with odd and even number of
particles, respectively. In Fig. \ref{spinb} we show the behavior of the
spin projection for an even number of particles and the difference between
the ground state energies for odd and even number of particles $E(B)$ as a
function of the magnetic field. For $B=0$, $E(0)=E$ and $\langle
g_{e}|S_{R}^{z}|g_{e}\rangle =0$, consistent with a time-reversal invariant
ground state.  In general $-1/4\leq \langle g_{e}|S_{R}^{z}|g_{e}\rangle
\leq 1/4$ and $0\leq E(B)=E_{\uparrow }\leq E$, and for $\Delta _{Z}\gg E$, 
$E(B)\rightarrow 0$ and $\langle g_{e}|S_{R}^{z}|g_{e}\rangle \rightarrow 1/4$

\section{Summary and discussion}

\label{sum}

We have calculated the low-energy eigenstates of a finite chain of a time-reversal-invariant topological 
superconductor numerically and in the particular case of an electron-hole symmetric band ($\mu=0$) 
also analytically. The analytical solution allows one to gain insight on the main features of the Majorana
zero-energy excitations at the ends of the topological chain and how the end states mix in the finite chain 
giving rise to low-energy excitations with finite energy. 

Using these solutions we have calculated the spin projection for each site $i$ of the chain $\langle S_{i}^{z}\rangle$
along the Rashba direction $z$ in finite chains. 
We show that excitations with
total spin projection $S_z= \pm 1/2$ fractionalize in two pieces with $S_e^z= \pm 1/4$ 
localized at each end of the chain. 
$\langle S_{i}^{z}\rangle$ displays oscillations with an 
exponential envelope. The decay length
of $\langle S_{i}^{z}\rangle$ at each end can be calculated solving a quartic equation with complex coefficients
for any $\mu$ and is well approximated by a simple analytical formula [Eq. (\ref{longe})].

Although we presented results for a specific model Hamiltonian, all the physical behavior discussed in the 
present work is generic of TRITOPS wires.

The finite energy excitation $E$ in chains with an odd number of particles with respect
to those with even number of particles should be experimentally detectable. In ordinary superconductors, 
this energy is of the order of the superconducting gap and has been measured in experiments in small islands in which 
capacitance effects allow researchers to control the number of electrons in small superconducting 
systems.\cite{lafar} 
Furthermore, these experiments permit to tune the chemical potential $\mu$ changing the localization length
of the states at the end of the chain and the excitation energy $E$ of the quasiparticles
entangling both ends.
In the present case, $E$ lies deep inside the superconducting gap. 
Alternatively, scanning tunneling microscope measurements  \cite{yazdani, pascual, franke} could also detect 
these subgap excitations.  

An application of a magnetic field opens other interesting possibilities.
The excitation energy $E$ is split by the magnetic field. Applying the field 
only to one end of the chain would open the possibility of analyzing the response of the fractional 
spin projections at the ends.
We identify two possible scenarios, depending on the amplitude
of the Zeeman splitting $\Delta_Z$ relative to the energy of the excitations without magnetic field $E$. 
In the case of $\Delta_Z \gg E$, which can be easily achieved for
very long chains, assuming that the field is applied to right end favoring spin up there, 
the ground state for an even number of particles has expectation value of
the spin projection at the left in the range $-1/4 \leq S_{L}^{z} \leq 0$ 
and $0 \leq S_{R}^{z} \leq 1/4$ at the right,
while the lowest two eigenstates with odd number of particles have 
$S_{L}^{z}=S_{R}^{z}= \pm 1/4$. Then,  the one-particle excitation energies correspond 
to flip a fractional spin at the left without energy cost, or flipping it at the right
with an energy cost $\Delta_Z/2$, the usual one for creating a spin down.

On the other hand, for $\Delta_Z \ll E$,
the entanglement between left and right end
excitations manifests itself in the magnetic response. 
In this case, the ground state for an even number of particles has 
$S_{L}^{z}=S_{R}^{z}=0$, while for odd number of particles still $S_{L}^{z}=S_{R}^{z}= \pm 1/4$.
Clearly, there is a Zeeman splitting of the one-particle excitations equal to $\Delta_Z/2$. 
This is precisely  half the magnitude of the one
expected for a ordinary spin $1/2$ and reflects the fact that $S_z$ is fractionalized, 
with $S_e^z= \pm 1/4$ at the ends.

Scanning tunneling microscope measurements akin to those used to 
investigate Yu-Shiba-Rusinov excitations induced by magnetic impurities should be able to detect these features 
in the subgap spectrum of TRITOPS wires. \cite{yazdani, pascual, franke}
We believe that also microwave radiation \cite{tosi,hays} can produce transitions between the quasiparticles split
by $\Delta_Z/2$ for a small magnetic field.

\section*{Acknowledgments}

A. A. A. is sponsored by PIP 112-201501-00506 of CONICET and PICT 2013-1045
of the ANPCyT. We acknowledge support from CONICET, and UBACyT, Argentina  and  the Alexander von Humboldt
Foundation, Germany.

\appendix
\section{Summary of the method by Alase et al in the Namb\'u formalism}
We start by expressing  the Hamiltonian of Eq. (\ref{ham})  in terms of
 Nambu spinors $\psi_j^{\dagger}= \left( c^{\dagger}_{j \uparrow}, \; c^{\dagger}_{j \downarrow}, \; c_{j \downarrow},\; -c_{j \uparrow} \right)$. The result is
 \begin{equation}
H=\sum_{j=1}^L \psi_j^{\dagger} h_0 \psi_j + \sum_{j=1}^{L-1} \left( \psi_j^{\dagger} h_1 \psi_{j+1}  + H. c. \right),
\end{equation}
with
 \begin{eqnarray}
h_0 &=&  \frac{\mu}{2} \sigma_0 \otimes \tau_z, \nonumber \\
h_1 &= & - \frac{t}{2} \sigma_0 \otimes \tau_z + i \frac{\lambda}{2}  \sigma_z \otimes \tau_z + \Delta \tau_x.
\end{eqnarray}
Here $\sigma_j$ and $\tau_j$ are Pauli matrices acting on the spin and particle-hole degrees of freedom, respectively, while $\sigma_0$ and $\tau_0$ are $2 \times 2$ unit matrices.

 We define the state $|j \rangle $  associated to the Nambu operator $\psi_j$ such that
\begin{equation}
H=\sum_{j=1}^L h_0 |j \rangle \langle j | + \sum_{j=1}^{L-1} \left( h_1 | j \rangle \langle j+1 | + H. c. \right).
\end{equation}
In this notation we define projector operators over bulk ($P_B$) and boundary ($P_{\rm bou}$) as follows
\begin{equation}
P_B= \sum_{j=2}^{L-1} |j \rangle \langle j |, \;\;\;\;\;\;\;\: P_{\rm bou}=  |1 \rangle \langle 1 | + |L \rangle \langle L |.
\end{equation}
The projector $P_B$ is over all the sites in which all the Hamiltonian matrix elements are contained in the chain, while  $P_{\rm bou}$ contains the sites at the left and right ends of the chain.
They satisfy $P_B+P_{\rm bou}=1$. 

Following Alase et all, we  aim to solve the bulk-boundary
eigenvalue problem 
\begin{equation}
P_B H  |\Psi \rangle = E P_B |\Psi \rangle, \;\;\;\;\;\;\;  P_{\rm bou} H |\Psi \rangle = E P_{\rm bou} |\Psi \rangle.
\end{equation}
We construct a generalized Bloch state expanding in  powers of a complex number $z$ as follows
\begin{equation}
|\psi^{B}(z) \rangle= w_z  \sum_{j=1}^L z^{j-1} | j \rangle.
\end{equation}
The latter is represented with a spinor of the form
$| \psi^{B}(z) \rangle =\left(u_{\uparrow}(z)  , \;u_{\downarrow}(z) ,\;   v_{\downarrow}(z)  , \;-  v_{\uparrow}(z)  \right)^t$.
The coefficients $\left(u_{\sigma}(z),\; v_{\sigma}(z) \right)$ are determined in order to satisfy 
\begin{equation} \label{bulk}
P_B \left[H-E(z) \right] | \psi^{B}(z) \rangle =0.
\end{equation}
The rest of the calculation continues in Eq. (\ref{eig}) to Eq. (\ref{e2}). The four eigenstates corresponding to the solution of the bulk problem 
(\ref{bulk})  are Nambu states 
 $| \psi^{B}_k \rangle \equiv |\psi^{B}(z_k) \rangle$
 of the bulk Hamiltonian with $
u_{\sigma, k}>0$ given by Eq. (\ref{vsu}) for $\sigma = \uparrow$ and the same equation
changing the sign of both $\Delta $ and $\lambda $
for $\sigma = \downarrow$

The last step is to find the solution of the full eigenvalue equation $(H-E) | \Psi \rangle =0$. To this end we
express the state $ | \Psi \rangle $ as a linear combination of the bulk eigenstates  $| \psi^B_k \rangle $,
\begin{equation}
| \Psi \rangle = \sum \alpha _{k} | \psi^B_k \rangle.
\end{equation} 
The coefficients $
\alpha _{k}$ and the energy $E$ are determined in order to satisfy
\begin{equation}
P_{\rm bou} \left[ H -E \right]  |\Psi \rangle =0.
\end{equation}
The corresponding coefficients are given by  Eqs. (\ref{ep1}) and (\ref{ep2}).


\begin{thebibliography}{9}
\bibitem{note} The terms proportional to $\alpha_3$ in Eqs.  (\ref{c1}) and (\ref{c2})
are proportional to $\bar{z}_{1}^{L-2}+(-\bar{z}_{1})^{L-2}$.
For odd $L$ they vanish and the excitations for both ends continue to be decoupled with $E=0$.
\end{thebibliography}
\end{document}